 \documentstyle[twocolumn,11pt]{article}

\begin{document}

\def\be{\begin{equation}}
\def\en{\end{equation}}
\def\bear{\begin{eqnarray}}
\def\enar{\end{eqnarray}}
\newcommand{\mn}{\mu\nu}
\newcommand{\rs}{\rho\sigma}
\newcommand{\square}
{\kern1pt\vbox{\hrule height 1.2pt\hbox{\vrule width1.2pt\hskip 3pt
\vbox{\vskip 6pt}\hskip 3pt\vrule width 0.6pt}\hrule height 0.6pt}
\kern1pt}

\begin{flushright}
WU-AP/34/93 \\
gr-qc/9402-~~~\\
February 9, 1994
\end{flushright}
\begin{center}
{\bf
\vskip 0.25cm
{\Large Generality of Inflation in a Planar Universe }}\\
(Revised version)
\vskip 0.8cm

{\sc  Hisa-aki Shinkai}$^{1}$
{}~and
{\sc Kei-ichi Maeda}$^{2}$
\vskip 0.8cm

 Department of Physics, Waseda
University,\\ Shinjuku, Tokyo 169-50,  Japan
\vskip 0.5cm

{\bf abstract}
\vskip 0.15cm

\begin{minipage}[c]{9.5cm}
\small

We study a generality of an inflationary scenario by
integrating the Einstein equations numerically in a
plane-symmetric spacetime. We consider the inhomogeneous
spacetimes due to (i)
 localized gravitational waves  with a positive
cosmological  constant $\Lambda$, and (ii) an inhomogeneous
inflaton field $\Phi$ with a potential $\frac12 m^2 \Phi^2$.
For the case
(i), we find that any initial inhomogeneities are
smoothed out even if waves collide,
 so that we conclude that inhomogeneity due to
gravitational
waves do not prevent the onset of inflation. As for the case
(ii), if the  mean value of the inflaton field is initially
as large as the condition in an isotropic and homogeneous inflationary
model (i.e., the  mean value is larger than several times Planck mass),
the field is soon homogenized and the universe always evolves into de
Sitter spacetime. These  support  the cosmic no hair conjecture in a
planar universe.
We also discuss the effects of an additional massless scalar field,
which is introduced to set initial data in usual analysis.\\

\noindent
\small
Key Words: \parbox[t]{7.5cm}
{ \small General Relativity,
Cosmology, \\ Inflationary Universe.}

\end{minipage}
\end{center}
\vspace{0.5cm }

\noindent
{\footnotesize 1~ Electronic address:
62L508@jpnwas00.bitnet  ~or~
62L508@cfi.waseda.ac.jp }

\noindent
{\footnotesize 2~ Electronic address: maeda@jpnwas00.bitnet ~or~
maeda@cfi.waseda.ac.jp\\
 }

\small
\baselineskip = 15pt
\section{Introduction}

Over a decade, the inflationary universe model \cite{inflation}
is widely accepted as a standard scenario in the history of the early
universe.  Although
the fundamental idea of inflation is very simple, this explains many
difficulties in the standard big-bang model.  One of the main subjects
for the current researches  is to study the generality and universality
of the inflationary model
itself.   That is, to investigate the model
in anisotropic and/or inhomogeneous
spacetime, and to judge whether or not inflation
occurs naturally  for such a generic
spacetime.  This is, in other words, to study whether or not
the cosmic no hair conjecture \cite{nohair} is certified.

The rapid expansion in the inflationary era is caused by a  vacuum
energy described by an inflaton field, or equivalently by a positive
cosmological constant $\Lambda$.
For spacetimes with  $\Lambda$, the so-called cosmic no hair
conjecture has been proposed.
The conjecture is:  All initially expanding universes
with $\Lambda$ approach the de Sitter spacetime asymptotically.
The de Sitter spacetime is stable against linear perturbations
\cite{perturb}
and many models also support this conjecture\cite{nohair}.
There, however,
exist some counter examples \cite{counter-ex}.
So that we expect we may prove this conjecture with some
additional constraints.

For homogeneous but anisotropic spacetime, Wald \cite{wald}
showed that initially expanding spacetimes
(except Bianchi IX) with positive
$\Lambda$ approach the de Sitter spacetime within one Hubble
expansion time $\tau_H=(\Lambda / 3)^{-1/2}$.  And for Bianchi IX, the
additional initial condition that $\Lambda$
be larger than one third of the
maximum value of 3-dimensional Ricci scalar leads to the
same conclusion \cite{wald,kitamae}.
For inhomogeneous spacetimes, only one practical method we have
at present is to solve the Einstein equations numerically.

Assuming a plane symmetric spacetime, Kurki-Suonio, Centrella,
 Matzner and Wilson \cite{suonio} studied a phase transition
induced by the inflaton field
under the existence of initial thermal
fluctuations using new inflationary
model.  They found that inflation occurs only when the potential
is flat enough.  Since their motivation was different from studying
the cosmic no hair conjecture,
they did not consider the case which include large inhomogeneities
of spacetime.
Although they found de Sitter space is a final state, their initial
inhomogeneity is not enough large
to study the cosmic no hair conjecture.

Assuming a spherically symmetric spacetime, Goldwirth and Piran (GP)
studied the  behavior of inhomogeneous
distributions of scalar field.
{}From an estimation of an energy balance
between potential energy and scalar field
gradient term, they proposed a
criterion for the onset of inflation such that \be {{\rm
a~scale~length~of~inhomogeneity} \over
{\rm horizon~scale}} > {\rm a~few~times~} {\delta
\phi \over m_{pl}} \label{criterion} \en where $\delta \phi$ is a
spatial deviation of the scalar field from the
mean value and $m_{pl}$ is Planck mass.  They found that a
short scale inhomogeneity  compared to the horizon
scale leads spacetime  to collapse in closed
universe, and concluded that (\ref{criterion}) is sufficient
condition for chaotic inflationary models.
GP's simulations, however, are done under assumptions
 on the isocurvature initial data
with additional massless scalar field, which may cause homogeneous
expansion of the spacetime in the very initial stage. This
additional scalar field may change a homogenization process of
spacetime as we will see later.

In this paper, we study the validity of the cosmic no hair conjecture
in a plane symmetric spacetime. We have mainly two reasons
why we assume a planar
universe.  First, we investigate the behaviors of
the inhomogeneities due
to the gravitational waves, which cannot exist
in a spherically symmetric
spacetime.  It is well studied that any linear gravitational
waves  decay and disappear immediately
in de Sitter spacetime  \cite{weakgw}, so our
main interests are the spacetimes with strong gravitational waves.
As some numerical study in flat spacetimes shows,
if the strong gravitational waves are localized, those come to have
apparent horizons and collapse into  black holes \cite{stronggw}.
If such a strong gravitational field is  localized in a plane
symmetric spacetime, then  the spacetime  may evolve into a naked
singularity.
In fact,  Szekeres and Khan-Penrose \cite{colwave}
found exact solutions of colliding plane waves
in Minkowski background spacetime, which form a singularity.
If the spacetime is expanding, on the other hand,
the cosmological expansion forces to extend localized waves,
which competes to the above attractive force
due to nonlinear effect of gravity.
One of our purposes is to compare the behavior of those
competitive two effects.
Since the attractive force acts also for a localized
inhomogeneity due to an inflaton, we also study
a inhomogeneous inflaton in a plane symmetrical spacetime and
compare the effects to those in spherical symmetric spacetime
studied by GP.

Secondly, we are interested in the formations of
a naked singularity, since a singularity formed in a plane symmetric
spacetime is always naked.  Recently,
Nakao, Maeda, Nakamura and Oohara examined time symmetrical initial
data for Brill waves in an axial symmetrical spacetime with
cosmological constant and found that no waves with large
gravitational mass provide trapped surfaces \cite{nakaogw}.
Nakao also found that a dust sphere with
large gravitational mass in a background de Sitter spacetime
does not collapse into a black hole spacetime \cite{nakaodust}.
And it is also shown
that there exists an upper bound on the area of the apparent horizon
of a black hole  in an asymptotically de Sitter
spacetime \cite{shiromizu}. So that we may
conjecture that a large inhomogeneity
does not necessarily  prevent cosmological constant-driven inflation.
This conjecture has been also confirmed by the simulation of
non-rotating-axisymmetric vacuum spacetime with cosmological constant
\cite{shibata}.
{}From these results, it seems plausible that in an inflationary era,
inhomogeneities will simply evolve into many
small black holes in a background de
Sitter universe; the only worrisome possibility
is that a naked singularity
might form.  Disposing of this worrisome possibility  may thus
become the main problem in the study of
the cosmic no hair conjecture. Our
1-dimensional case seems well suited to address this problem.

In Section 2, we describe basic equations and our numerical procedures.
We show the effects of inhomogeneities due to
gravitational waves in Section 3, and those due to inflaton field
in Section 4. Conclusions and remarks are in Section 5.
As units, we use $8 \pi G =1$ and  $c=1$.

\section{Equations and Numerical Methods}

\subsection{The models}

We consider following two
effects on the evolution of the spacetime;
[I] inhomogeneities due to gravitational waves and
[II] inhomogeneities due to an inflaton field.
To make the problem simple, we set models as follows:

\begin{enumerate}
\renewcommand{\labelenumi}{[\Roman{enumi}]}
\item
Initial large inhomogeneities due to gravitational waves exist  in the
spacetime with positive cosmological constant $\Lambda$ (no inflaton
field). We are interested in the competitive processes
 between the expansion of the background spacetime and
the attractive forces of the gravity.

\item
Introduce an inhomogeneous inflaton field $\Phi$ in the context of
the chaotic inflationary model.
In a realistic inflationary model, it is natural to assume such
an inhomogeneous inflaton field.
Our interest is the effects of the spatial gradient
term of the field on the evolution of the spacetime,
which may prevent the onset of the inflation.
\end{enumerate}

\noindent

For the second model, we make initial data in two ways.
One is the similar setting of initial data as those of GP's,
i.e., the spacetime
with another massless scalar field $\Psi$ to obtain an isocurvature
initially. The other is without such an additional scalar field.
We examine the contributions of $\Psi$ to evolutions of the
spacetime.

In the chaotic inflationary scenario,
we usually assume that the
average of the inflaton,
$\bar{\Phi}$, is larger than {\it a few }$m_{pl}$ in order
for the Universe to expand sufficiently.
Even under an existence of  large inflaton field as an
average ($\bar{\Phi} > ~a~few~m_{pl}$),
it is not trivial how the large inhomogeneity of $\Phi$
contributes to the evolution of the Universe.
We analyzed
whether or not
such an gradient term in the energy $\rho_{\Phi'}$  prevents from
the onset of inflation.

\subsection{Field Equations}

We use the Arnowitt-Deser-Misner (ADM) formalism  \cite{adm1} to
 solve the Einstein equation
\be
R_{\mu \nu}-{1 \over 2} g_{\mu \nu} R+\Lambda g_{\mu \nu} = T_{\mu
\nu}, \label{einstein} \en
with the metric
\be
ds^2 = -(\alpha^2 -\beta^2 /\gamma_{11})dt^2+2\beta dtdx
+\gamma_{11}dx^2+\gamma_{22}dy^2+\gamma_{23}dydz+\gamma_{33}dz^2,
\label{metric0}  \label{metric}
\en
where the lapse function
$\alpha$, the shift vector $\beta_i=(\beta,0,0)$, and the 3-metric
$\gamma_{ij}$  depend  only on
time $t$ and one spatial coordinate $x$.
The extrinsic curvature $K^i_{~j}$ becomes in a matrix form
\be
K^i_{~j}=\left( \begin{array}{ccc}
                 K^1_{~1} & 0 & 0 \\
                 0 & K^2_{~2} & K^2_{~3} \\
                 0 & K^3_{~2} & K^3_{~3} \end{array} \right) .
\label{kmetric}
\en

Following  ADM formalism, Hamiltonian and momentum
constraint equations in our metric are written as
\bear
{}~^{(3)\!}R+({\rm tr}K)^2-K_{ij}K^{ij}&=&2  \rho_H + 2\Lambda,
\label{hamiltco} \\
D_1 ( K^1_{~1}- {\rm tr}K)&=&  J_1,
\label{momentco}
\enar
and  evolution equations are
\bear
{\partial_t} \gamma_{ij} &=& -2\alpha
K_{ij}+D_j\beta_{i}+D_i\beta_{j},
 \label{hatten1}  \\
{\partial_t} K_{ij} &=& \alpha
(~^{(3)\!}R_{ij}+{\rm tr}KK_{ij})-2\alpha K_{il}K^l_{~j} -D_iD_j\alpha
\nonumber \\ &~&+(D_j\beta^m)K_{mi}+(D_i\beta^m)K_{mj}+\beta^m D_m
K_{ij}-\alpha \gamma_{ij}\Lambda \nonumber \\
&~& - \alpha \{ S_{ij}+{1 \over 2} \gamma_{ij}(\rho_H-{\rm
tr}S)\}, \label{hatten2}  \end{eqnarray}
where Greek indices $i,j,k,...$ move 1 to 3, $D_i$ is a covariant
derivative on the 3-dimensional
hypersurface $\Sigma$ determined by $\gamma_{ij}$,
$~^{(3)\!}R_{ij}$ is 3-dimensional Ricci tensor  and $~^{(3)\!}R$ is
its Ricci scalar. $\rho_H$, $J_i$ and $S_{ij}$ are energy density,
momentum density and stress tensor,
respectively, defined by a observer  moving along
$n_\mu=(-\alpha,0,0,0)$. They are given as
 \be
\rho_H \equiv  T_{\mu \nu}n^{\mu}n^{\nu},~~
J_i \equiv  -T_{\mu \nu}h^{~\mu}_in^{\nu},~~
S_{ij} \equiv T_{\mu \nu}h^{~\mu}_ih^{~\nu}_j
\en
where $h_{~\mu}^{\nu}=\delta_{~\mu}^{\nu}+n^\nu n_\mu$ is
 the projection operator onto $\Sigma$.

In section 3, we will treat vacuum spacetime, i.e. $T_{\mn}=0$,
under the assumption of the existence of a positive cosmological
 constant $\Lambda$, while in
section 4, we will treat a scalar
field $\Phi$ with potential $V(\Phi)={1 \over 2} m^2 \Phi^2$ as a
source of the inflation  rather than $\Lambda$.
The energy momentum tensor of the scalar field is given by
\be
T_{\mn}=\partial_\mu \Phi \partial_\nu \Phi- g_{\mn}[{1 \over
2}(\nabla \Phi)^2 + V(\Phi)]. \label{tmunu}
\en
The scalar field evolution, then,  obeys the  Klein-Gordon
equation  \be
 \square \Phi ={dV \over d\Phi}. \label{sf-eq}
\en
To integrate (\ref{sf-eq}), we introduce the conjugate momentum
\be
 \Pi ={\sqrt{\gamma} \over \alpha} (-\partial_t {\Phi}+{\beta
 \over \gamma_{11}} \partial_x \Phi),
 \en
where
$\gamma = \det \gamma_{ij}$, and write down eq.(\ref{sf-eq})
into  two first-order partial
differential equations:
\bear \partial_t \Phi &=&{\beta \over \gamma_{11}}
\partial_x \Phi-{\alpha \over \sqrt{\gamma}}\Pi, \label{sc1} \\
\partial_t \Pi &=& \alpha \sqrt{\gamma}{dV \over d\Phi} +
\partial_x {1 \over \gamma_{11}}
 [\beta  \Pi -\alpha \sqrt{\gamma} \partial_x\Phi ]. \label{sc2}
\enar
The dynamical variables
are $\gamma_{ij}$ and $ K_{ij}$ (and  $\Phi$ and $\Pi$, when a
scalar field exists).

We impose periodic boundary condition,
 i.e., $f(x_1,t) \equiv
f(x_{\rm N+1},t)$, where $N$ is the grid number in $x$-direction.

Our simulation procedure follows Nakamura, Maeda, Miyama and Sasaki
\cite{adm3}:
\begin{enumerate}
\renewcommand{\labelenumi}{(\roman{enumi})}
\item
 Determine initial values by solving two constraint equations.
\item
 Evolve time slices by using the  dynamical equations.
\item
 Check the results of (i) and (ii)  using the two constraint
equations on every time slice.
\end{enumerate}
\noindent
Explicit procedure is described in the section 2.4.

So far, the Texas group \cite{ACM} has constructed  a numerical
code for a planar cosmology. Their metric is different from ours
 (\ref{metric}). They adopt a
diagonalized spatial metric, which fixes the shift vector.
Their time developing procedure  is also different from ours.
They use both constraint
and dynamical equations to get the stable solutions.


\subsection{Initial Value Problem}

 We use the  York-O'Murchadha's conformal approach \cite{initial0}
to get
initial values. Defining conformal factor $\phi(x)$ and setting
\bear
\gamma_{ij} &=& \phi^4 \hat{\gamma}_{ij},  \label{initialconf} \\
K_{ij} &=& \phi^{-2}\hat{K}_{ij}^{TF}+
{1 \over 3}\phi^4\hat{\gamma}_{ij}  \mbox{tr} K
\\ \rho &=& \phi^{-6} \hat{\rho},
{}~~~J^i = \phi^{-10} \hat{J}^i,
\enar
where $\hat{K}_{ij}^{TF}$ is the  trace-free part of the extrinsic
curvature
$\hat{K}_{ij}$, and $\mbox{tr} \hat{K}(=\mbox{tr} K)$ is a trace
 part of $\hat{K}_{ij}$.
The
quantities with a caret denote physical variables
in the conformal frame.

The Hamiltonian and momentum constraint equations then become
\bear
{}~8~^{(3)\!}\hat{\Delta}
\phi&=&~^{(3)\!}\hat{R}\phi-\hat{K}_{ij}^{TF}
\hat{K}^{ij}_{TF}\phi^{-7} +2[{1 \over 3}
(\mbox{tr} \hat{K})^2 - \Lambda -\rho]\phi^5
\label{inithamilt} \\
\hat{D}_j\hat{K}^{ij}_{TF}&=&{2 \over 3}\phi^6\hat{D}^i \mbox{tr}
\hat{K} +\hat{J}^i \label{initmoment}
\enar
where $^{(3)\!}\hat{\Delta}$ is
the 3-dimensional Laplacian defined by $\hat{\gamma}_{ij}$,
$~^{(3)\!}\hat{R}$ is  the scalar curvature.
In this approach, $\hat{\gamma}_{ij}, \mbox{tr}\hat{K}$ and the
transverse-traceless(TT) part of the extrinsic curvature
$\hat{K}^{TT}_{ij}$ (and $\hat{\rho}$ and $\hat{J}_i$ when we
 include matter) are  left to our choice.

We impose constant-mean-curvature condition
$\mbox{tr}\hat{K}={\rm constant.}$
 on the initial hypersurface. We also  assume that
we do not have momentum density ($\hat{J}_i=0$).
Then, for the initial data with  $K_{ij}^{TF}=0$,
the momentum constraint equation (\ref{initmoment})
becomes trivial.  Even in the case of $K_{ij}^{TF}\neq 0$,
we just have to give an arbitrary $K_{ij}^{TT}$.
We can assume that the longitudinal part of $K_{ij}^{TF}$
vanishes.

In order to get the periodic solution of (\ref{inithamilt}),
the value of $\mbox{tr}\hat{K}$ which we give on the initial
hypersurface should be chosen so that such a solution exists.
The condition for $\mbox{tr}\hat{K}$ depends on the models.
 Details are shown in the following sections.
We use the 5th order Runge-Kutta method (Fehlberg method) to solve
$\phi$ in (\ref{inithamilt}).

\subsection{Time Developing Scheme}

For coordinate conditions, we impose a geodesic slicing
condition such as $\alpha =1$ and $\beta=0$.
This slicing condition entails the risk
 that our numerical hypersurface may hit a singularity
and  stop there. If  no singularity appears, however,  then it may be
the best coordinate condition  for revealing whether de Sitter space
emerges as a result of time evolution.

We use finite differential scheme to integrate the Einstein
equations (\ref{hatten1}) and (\ref{hatten2}) with 400
grids in the spatial
direction $x$.  We use a simple central finite difference
with second order
scheme to determine derivatives.

The time step $\Delta t$ on each slice is determined, as
to keep the accuracy of the computations.
We set $\Delta t$ by imposing that a volume element $\gamma$
and a expansion rate tr$K$ do not change too much
($\pm 0.1\%$ change) in each step.

The accuracy is checked by comparing both sides
of constraint equations.
  For Hamiltonian constraint equation in the case of
vacuum, for example, we calculate \be
Err(t,x)={(R^{(3)}+K^2)-(K_{ij}K^{ij}+2 \Lambda +2\rho_H) \over
 |R^{(3)}|+K^2+K_{ij}K^{ij}+2 \Lambda+|2\rho_H|}
\en
at the point $x$ on $\Sigma (t)$
and regard its maximum value
$Error(t)={\rm max} \{ Err(t,x) | { ~x \in \Sigma (t) }\}$
on each hypersurface as the error on that slice.
A similar
definition is performed for momentum constraint equation.
An example of $Error(t)$ is shown in Fig. 2.1.
In our calculation, the
maximum error in the Hamiltonian constraint
equation is a few times of $O(10^{-4})$ on the initial hypersurface,
 and this accuracy is maintained even after time evolutions.

Our overall procedures to evolve the system are as follows.
At first stage, we solve the initial value on a hypersurface
$\Sigma_1$, and write $\gamma_{ij}$ and $K_{ij}$ as
$\gamma_{ij}(1)$ and $K_{ij}(1)$.

\begin{enumerate}
\item Evaluate the time step $\Delta t$.
\item Define next hypersurface $\Sigma_2$
and $\Sigma_3$ at $t=t+\Delta
t/2$ and $t=t+\Delta t$, respectively.
\item Evolve the metric on $\Sigma_2$ ($\gamma_{ij}(2)$ and
$K_{ij}(2)$), using (\ref{hatten1}) and (\ref{hatten2}) on $\Sigma_1$.
\item If the scalar field exists, evolve them onto $\Sigma_2$
($\Phi(2)$ and $\Pi_\Phi(2)$), using (\ref{sc1}) and
(\ref{sc2}) on $\Sigma_1$.
\item If the massless scalar field exists (in the case of isocurvature
model in section 4.1), evolve them onto $\Sigma_2$
($\Psi(2)$ and $\Pi_\Psi(2)$), using (\ref{sc1}) and
(\ref{sc2}) for $\Psi$ on $\Sigma_1$.
\item Calculate r.h.s. of (\ref{hatten1}), (\ref{hatten2}),
(\ref{sc1}) and (\ref{sc2}) on $\Sigma_2$.
\item Evolve values on $\Sigma_3$, using leap-frog method.
\item Check the increasing ratios of $\gamma$ and $K$. If one of
them exceeds a certain value ($\pm 0.1 \%$ change),
then control $\Delta t$ and return to {\rm 2.} again.
\item Check the accuracy of the computations by constraint equations.
\item Replace the values on $\Sigma_3$ by those on $\Sigma_1$, and
turn to 1.
\end{enumerate}


\section{Inhomogeneities due to gravitational waves}

In this section, we show the evolutions of the inhomogeneities due
to gravitational waves in the spacetime with positive cosmological
constant $\Lambda$. We expect two competitive
effects: one is the expansion of space
 due to cosmological constant, the
other is the attractive forces due to the nonlinearity of the
gravity.  We examine whether  such a spacetime leads to an inflationary
era, and whether such  initial inhomogeneities and anisotropies smooth
out during inflation periods.

As we already mentioned in the introduction, linear gravitational
waves are always decay in the de Sitter background spacetime.
This result is also confirmed in our numerical calculations as
one of our code tests. We, therefore, concentrate
our attentions to the large inhomogeneities
due to gravitational waves. The summary
 has been reported in \cite{WUAP29}.

Since we have  $\Lambda$ in this system, we adopt the Hubble
expansion time $\tau_H=(\Lambda / 3)^{-1/2}$
 as our time unit, which is a characteristic expansion time of the
universe.  Our unit of length is also normalized to the
horizon length of the de Sitter spacetime $l_H=(\Lambda / 3)^{-1/2}$.

To evaluate the inhomogeneities on each hypersurfaces, we use three
different curvature invariants. First one is the 3-dimensional
Riemann invariant scalar  $~^{(3)\!}R_{ijkl}~^{(3)\!}R^{ijkl}$,
where  $~^{(3)\!}R_{ijkl}$ is the Riemann tensor of the 3-metric on
$\Sigma$. We use its dimensionless value normalized by
the  cosmological  constant,
 \be
{\cal C}(t,x)\equiv { \sqrt{~^{(3)\!}R_{ijkl}~^{(3)\!}R^{ijkl}}
\over \Lambda}~~~~~{\rm on} ~\Sigma(t), \label{curvature}
\en
and call it  the  ``curvature" hereafter.   We estimate the
magnitude  of the inhomogeneities in the 3-space $\Sigma$ by the
maximum value of this ``curvature"  on each slice,  i.e. ${\cal
C}_{\rm max}(t) = \max\{ {\cal C}(t,x)~| { ~x \in \Sigma (t) }\} $.

Second one  is an invariant scalar induced from the Weyl
tensor $C_{\mu \nu \rho \sigma}$.
For the vacuum spacetime, if the  Weyl tensor
vanishes and no singularity appears,
 the spacetime is  homogeneous and isotropic. Hence this can be used to
check whether de Sitter universe is recovered or not. We use the
decomposition of the Weyl tensor:
\be
E_{\rho \sigma}=~~C_{\rho\mu \sigma
\nu}n^\mu n^\nu , ~~~
B_{\rho \sigma}=~^\ast C_{\rho\mu \sigma \nu}n^\mu
n^\nu,
\en
where $~^\ast C_{\mu \nu \rho \sigma}\equiv {1 \over 2}C^{\alpha
\beta} _{~~\rho \sigma}  \varepsilon_{\alpha\beta\mu \nu}$ is the
dual of the Weyl tensor and $n^\mu$ is a timelike vector
 orthogonal to the
hypersurface $\Sigma$.  In  analogy to the electromagnetism, the
3-dimensional variables $E_{\rho \sigma}$ and  $B_{\rho \sigma}$ are
called an electric and a magnetic parts of the Weyl tensor. We
can  reconstruct the Weyl tensor completely from this pair of
tensors.  Moreover, those two variables are easy to drive out using
our dynamical variables \cite{smarr-ann}:
\bear
E_{\rho \sigma}&=&~^{(3)\!}R_{\rho \sigma}+KK_{\rho
\sigma}-K_\rho^{~\mu}K_{\sigma \mu}, \\
B_{\rho \sigma}&=&  \varepsilon_{\mu \nu (\rho} D^\mu
K^\nu_{~\sigma)} .\enar
To estimate the field's inhomogeneities, we introduce
\be {\cal H}(t,x) =E_{\mu
\nu}E^{\mu \nu}+B_{\mu \nu}B^{\mu \nu}, \label{superenergy}
\en
as a sort of gravitational ``super-energy". In fact, it is the purely
timelike component of the Bel-Robinson tensor \cite{belrob}. If
${\cal H}(t,x) \rightarrow 0$,  then  $C_{\mu \nu \rho \sigma}
\rightarrow 0$ unless the hypersurface becomes null.

Third invariant is 4-dimensional Riemann invariant scalar
$R_{\mu \nu \rho \sigma}R^{\mu \nu \rho \sigma}$. We calculate
this variable by using a relation between Riemann tensor and Weyl
tensor \be
R_{\mu \nu \rho \sigma}R^{\mu \nu \rho \sigma} =C_{\mu \nu \rho
\sigma}C^{\mu \nu \rho \sigma}+2R_{\mu \nu}R^{\mu \nu}-{1\over
3}R^2. \label{4curvature} \en  In the de Sitter case (isotropic and
homogeneous case),  $R_{\mu \nu \rho \sigma}R^{\mu \nu \rho \sigma}={8
\over 3}\Lambda^2$, since $C_{\mu \nu \rho
\sigma}=0,~~R_{\mu\nu}=\Lambda g_{\mu \nu}$ and $R=4\Lambda$.

The initial values are determined as follows.
We treat pure gravitational waves with following two cases:
{\bf [case 1]} The inhomogeneities reside in the 3-metric
$\hat{\gamma}_{ij}$ and  $\hat{K}_{ij}^{TT} =0$.
We set a pulselike distortion initially expressed by
\be {\rm diag} (\hat{\gamma}_{ij})=(1, 1+a~{\rm
e}^{-(x/x_0)^2},  1), \label{exp-type}\en
where $a$ and $x_0$ are free parameters.
{\bf [case 2]} The 3-metric is conformally flat and the
inhomogeneities reside in  $\hat{K}_{ij}^{TT}$, i.e.,
\be {\rm diag} (\hat{K}^{TT}_{ij}) =
\tilde{a}~{\rm e}^{-(x/\tilde{x}_0)^2}
( 0, 1, -1), \label{cos-type1}\en
 where $\tilde{a}$ and $\tilde{x}_0$ are also free parameters.
Initial data we investigated are listed in Table 3.1 and 3.2.
The results for {\bf [case 2]}
are quite similar to  those of those of {\bf [case 1]}.

In order to get a consistent data under periodic boundary condition,
we set $\hat{K}$ as
\be
\mbox{tr}\hat{K}=-\sqrt{3 \Lambda}(1+\delta_K)
\label{tracecon}
\en
on the initial hypersurface $\Sigma(t=0)$.
Here, $\delta_K$ is a
positive  constant, which should be introduced because the periodic
 pulse increases expansion rate of the universe besides $\Lambda$.
By integrating (\ref{inithamilt}), we get conformal factor $\phi$.
There is one free parameter $\lambda$, which fixes the scale. If $\phi$
is multiplied by $\lambda$, we find another solution by replacing
$\delta_K$ with $\delta_K / \lambda^2$, which is physically
the same as the previous solution.
Hence, if we  impose  $\phi \rightarrow 1$  at the numerical boundary,
then the value of $\delta_K$ is fixed; it is less than $O(10^{-2})$.

A pulse-like wave has two characteristic physical dimensions, a width
and an amplitude. We use ${\cal C}_{\rm  max}(t)$ as an amplitude
measure, and we define the width $l(t)$   by the proper distance
between two points where $\gamma$
(the square of the 3-volume) decreases by  half  from its maximum
value  $\gamma_{max}$. An example of the initial data is shown in
Fig. 3.1.  This is the case of $l(t=0)=0.086l_H, {\cal C}_{\rm
max}(t=0)=9.67$.

The behaviors of the time evolutions are following.
Initial distortion begins to propagate both  in the
$\pm x$ directions.  We see the initial curvature ${\cal C}(t=0,x)$
localized at
the center with three peaks [Fig. 3.1(b)]  separates just after $t=0$,
and propagate at the light speed.
In Fig. 3.2, we show both
``curvature" ${\cal C}(t,x)$ and ``super-energy" ${\cal H}(t,x)$.
Two waves moves to the numerical boundaries and collides
each other on the boundary, since we assume periodic boundary condition.
After the collision, two waves proceed again in their original
directions.
This behavior is reminiscent of solitonic waves.
We also show, in Fig. 3.3, ${\cal C}_{\rm max}(t)$
and the invariant ${\cal I}(t,x)=\sqrt{R_{\mu \nu \rho
\sigma}R^{\mu \nu \rho \sigma}}/\Lambda$.
We can see clearly that the collision of the waves occurs
around $t=0.2 \tau_H$, when  the ``curvature" is superposed.
At the final stage, the spacetime is homogenized
by the expansion of the universe.
We can see ${\cal I}(t,x) \rightarrow
\sqrt{8/3}$ (homogeneous de Sitter spacetime)
 within one Hubble expansion time $\tau_H$ as expected
[see eq.(\ref{4curvature})].
We see the spacetime finally
succumbs to the overall expansion driven by  the cosmological constant,
and becomes indistinguishable from de Sitter spacetime.

In our simulation,  $l$ and ${\cal C}_{\rm max}(0)$ range for
{\bf [case 1]}
between  $0.080 l_H \leq l \leq 2.5 l_H$ and
$0.020\leq{\cal C}_{\rm max}(0)
\leq 125.0$.
 We also computed the evolution of
{\bf [case 2]}, 
 and found that
all initial inhomogeneities decay and disappeared within
one Hubble expansion time.
 From the present results,  we would conclude that for any
large Riemann invariant and/or small width inhomogeneity on
the initial
hypersurface, the nonlinearity of the gravity has little effect
and the spacetime always  evolves into a de Sitter spacetime.

One of our motivation is to see whether or not the singularity occurs
 in the presence of a cosmological constant.
In the above simulation,  collisions occur at the
boundary, which may make some troubles.
In order to avoid such numerical difficulties, we also
 construct  initial data representing two nearby pulse waves in one
numerical range.
An example of the time evolution of the curvature
 has been shown in Figure 3.4.
We find again that all inhomogeneities  decay and disappeared.
 For a wide range of initial widths and ``curvatures"
($0.080 l_H \leq l \leq 0.10 l_H$, $40.0 \leq{\cal C}_{\rm
max}(0)\leq 125.0$ and the periodic distance $d$ is  $0.20l_H
\leq d \leq  0.50l_H$),  all inhomogeneities  decay below  1 \% of
their initial ``curvatures" within one Hubble expansion time.


The results that all the spacetimes we analyze
are homogenized even if we include collision of  waves  occurs first
are consistent with the calculations by
Centrella and Matzner \cite{cenmatz}.
They examined the collision of gravitational shock
waves in an expanding Kasner background both analytically and
numerically,
and concluded that such a collision leads to no  singularity.
The expansion
due to the cosmological constant in our simulation is exponential
--- much faster than Kasner (power-law) expansion.

One may wonder what happens in the limit of $\Lambda \rightarrow 0$
or $l \rightarrow 0$, when we expect no effect of the cosmological
constant.
We also consider the cases with no cosmological constant.
In a plane-symmetric spacetime,
if we impose periodic boundary condition, then we
should set tr$K \neq 0$ due to the contribution of the energy of
gravitational waves.
That is, the background spacetime  expand or contract.
The preliminary results of the time evolutions show  that such a
periodic wave propagate with decreasing its amplitude
[${\cal C}_{\rm max}(t) \sim \gamma^{-1/3}$] if the background is
expanding (before colliding to another waves).
This suggests that single gravitational wave in
 a plane-symmetric expanding spacetime do not collapse itself.
The evolutions for the case of colliding waves and for the case
in a  contracting background are still under studying.
The details will be reported in the elsewhere.

\section{Inhomogeneities due to scalar field}
In this section, we introduce a scalar field $\Phi(x,t)$ as a source of
inflation instead of cosmological constant,
and show the results of the behavior of its inhomogeneities to the
evolution of the spacetime. In the usual analytic approach of
 inflationary scenario, the inflaton
field is assumed    homogeneous. By introducing inhomogeneities, the
energy density of the scalar field $\rho_\Phi$ is written as
\bear
\rho_{\Phi}&=&{1 \over 2\gamma} \Pi^2 + {1 \over 2 \gamma_{11}}
(\partial_x \Phi)^2 + V(\Phi) \nonumber \\
&\equiv&\rho_{\dot{\Phi}} + \rho_{ \Phi'} + V(\Phi) ,
\label{rhodef}
\enar
where $V(\Phi)$ is the potential term of the scalar field.
Our interests is how the gradient term $\rho_{ \Phi'}$
 affects on the expansion of the universe.
The initial kinetic term $\rho_{\dot{\Phi}}$
gives less effects on the inflationary scenario.

 The same-aimed simulations have been already
done by GP  using a spherically symmetric code \cite{GP}, and one of
their results is that the spacetime with large $\rho_{ \Phi'}$
do not evolve into the inflationary era.
 In order to compare the results between
ours and  theirs, i.e., the difference of the symmetry of the
spacetime, we first set similar initial situations and parameters
with GP, that is with `isocurvature initial data'  (section 4.1).
This condition makes us easy to determine
initial data with constant mean curvature slicing, but also makes
us unclear the effects of inflaton $\Phi$ itself,
because the spacetime may expand homogeneously in the very
initial stage by introducing another scalar field  $\Psi$,
and such an expansion might reduce an initial large inhomogeneity
of inflaton field.
So that we also prepare the initial data without
 $\Psi$, and computated their
time evolutions (section 4.2).

\subsection{Evolution of the isocurvature initial data}

We study the evolutions of inhomogeneous scalar field with
isocurvature initial data following to GP. In addition to the inflaton
field, we introduce a massless scalar field $\Psi$ such that
the initial  total energy density
\be
\rho_{\rm total}=\rho_\Phi+\rho_\Psi
\en
 becomes uniform, where  $\rho_\Psi=\rho_{\dot{\Psi}}
+ \rho_{ \Psi'}$.
In this setting, the Hamiltonian constraint equation (\ref{hamiltco}) is
initially trivial under constant mean
curvature slicing (\ref{tracecon2}).
This method was proposed by GP. We think that it will
 help to compare our results with theirs to work with same initial
situations.
The massless field $\Psi$ is introduced as an additional radiation
field,
and is expected to dissipate immediately by the expansion of the
universe. As we show later in Fig. 4.5, only inflaton $\Phi$ effects
to the spacetime evolution at the late stage.
The initial spacetime, however, is set to have uniform
 expansion rate (since $\rho_{\rm total}$=constant.),
so that the initial-isocurvature setting may not produce
an exact model  for our
aims to search the effects of inhomogeneity.
As we show in the next sub-section, we also prepare a fully
inhomogeneous initial data and follow those time evolutions.

In order to define dimensionless variables,  we introduce
 the effective cosmological constant as
$\Lambda_{\rm eff} \equiv  \rho_{\rm total}$ at $t=0$ and
normalize time and scale by the units
$\tau_H$ and $l_H$, where
$\tau_H=l_H=({\Lambda_{\rm eff}/3})^{-1/2}$.
The potential $V(\Phi)$ is chaotic inflationary
type; $V(\Phi)={1 \over 2} m^2 \Phi^2 $ with $m^2=0.01$.

Initial scalar distribution is set as
\be
\Phi_{\rm initial}={\Phi_0}+\delta\Phi \cos(2\pi {x \over \lambda}),
\label{cos-type-scalar} \en
where ${\Phi_0}, \delta\Phi$ and $\lambda$ are parameters,
which indicate
mean value of the scalar field, amplitude of inhomogeneity and scale
length of inhomogeneity, respectively.
We set the initial spacetime is
conformally flat and has no gravitational waves
 ($\hat{\gamma}_{ij}=\delta_{ij}$,
$\hat{K}_{ij}^{TT}=0$),
 and choose $\hat{K}$ as
\be
 \hat{K}=-\sqrt{3 \Lambda_{\rm eff}}.
\label{tracecon2}
\en

We choose $\Phi_0$ several times of $m_{\rm pl}$ in order to get
`suitable' expansion of the Universe when $\delta \Phi=0$.
This is from a condition for inflation in isotropic and
homogeneous universe.
We are interested in effects of gradient term of the scalar
field $\rho_{ \Phi'}$ to the time evolution of spacetime,
especially large $\rho_{ \Phi'}$ compare to
$\rho_{\rm total}$.
We get large $\rho_{ \Phi'}$, if we choose small $\lambda$ and
large $\delta\Phi$.
However, the total energy density $\rho_{total}=\rho_{\Phi'}+V(\Phi)$
at the inflationary era is order of the Planck scale $m_{pl}^4$, so that
if we introduce `suitable' $\Phi_0$, then acceptable values of
$\lambda$ and $\delta\Phi$ are limited.
But we can get large enough
$\rho_{ \Phi'}$ compare to $\rho_{\rm total}$ as will be shown in Table.

First, we prepare initial data of the scale $\lambda \simeq l_H$ and the
amplitude up to
$\delta\Phi \sim 0.1 \Phi_0$ (model [II-1a $\sim$ 1c] in Table 4.1).
Corresponding ratio of
 $\rho_\nabla \equiv \rho_{ \Phi'}+\rho_{ \Psi'}$ to
$\rho_{\rm total}$ are also shown in Table 4.1.
Those model have large contributions of $\rho_\nabla$, we may regard that
gradient term is locally dominated.
As we can see from Fig. 4.1, in which the ways of evolutions
(model [II-1b]) is shown,
 all the inhomogeneous scalar field
are first homogenized and the field go to the slow-rolling phase along
to the ordinary chaotic inflationary scenario.
We find the spacetime expands sufficiently
in the region where the inflaton has large enough $\Phi_0$.
This  confirms the criterion
(\ref{criterion}) proposed in spherically symmetric case by GP.

Next, we prepare initial data of the small $\lambda$. In these cases,
we may
expect a formation of the collapse of the spacetime, as an example
shown by GP. Our case has planar symmetry, so if the collapse occurs,
then the naked singularity appears.  Hence from this calculation we
can also discuss the so-called cosmic censorship in the spacetime with
inflaton field.
We prepare data with the same $\delta\Phi$ and different
$\lambda$'s such as $\lambda=l_H, {1 \over 2} l_H,
 {1 \over 4} l_H$ and ${1 \over 6} l_H$, when
$\Phi_0=8.0 m_{\rm pl}$ and $\delta\Phi=0.06 m_{\rm pl}$
 (model [II-1d $\sim$ 1g] in Table 4.1).
In the last case ($\lambda={1 \over 6}
l_H$), $\rho_\nabla$ is locally 86\% of the $\rho_{\rm total}$,
we can regard gradient term is locally dominated again.

The typical time evolution of $\Phi$ is shown in Fig. 4.2
(model [II-1g]). The evolution is quite similar to Fig. 4.1.
That is, in all the case, the field  $\Phi$ is homogenized
and the spacetime expands sufficiently.

For various scale of the deviation $\lambda$ (model [II-1d $\sim$
1g], we show the
trajectories of the evolution at $x=0$ in the $\Phi$-$\overline{\Pi}$
  phase space, where $\overline{\Pi}=-\Pi/\gamma$ (Fig. 4.3).
At the beginning, the system
shows damped oscillation around the mean  value of the
field in the phase space,  and the
trajectory will soon approach that of the homogeneous
case.  At the very end, the scalar field begins damped
oscillation
around the origin and finally reach the potential minimum.

To compare the effects of $\rho_\nabla$ on the evolutions of a
scalar field, we show the trajectories for various $\lambda$ in Fig.
4.4(a).  The behaviors are almost same as in Fig. 4.3,
and the merging
points to the  homogeneous case are independent of the initial
gradient term,  which is agree with the spherically symmetric
case \cite{GP}.  The contributions of $\rho_\nabla$
 to the evolutions of a universe are drawn in Fig. 4.4(b).
Effects on the evolution of the spacetime are compared
 using the maximum value of $\gamma$ on each hypersurface.
In the figure, the homogeneous initial data (both mean value
$\Phi_{\rm initial}=\Phi_0=8.0m_{\rm pl}$ and maximum value
$\Phi_{\rm initial}=\Phi_0+\delta \Phi=8.06m_{\rm pl}$)
are also plotted with
solid line, which are almost piled up but less than the cases with
inhomogeneous initial data.
So we can judge  $\rho_\nabla$ enhances the evolution of the
universe.
In Fig. 4.5, we show the proportions of the each terms
in energy density (
$\rho_{\dot{\Phi}}$, $\rho_{\Phi'}$,
$\rho_{\dot{\Psi}}$, $\rho_{\Psi'}$,
and $V(\Phi)$ )
to $\rho_{\rm total}$.
We see  that $\rho_{\nabla}=\rho_{\Phi'}+\rho_{\Psi'}$
becomes less effect within 1.0$\tau_H$ [Fig. 4.5(a)],
and the kinetic energy of the scalar field dominates
at the last stage [Fig. 4.5(b)].
The kinetic energy of the massless field $\rho_{\dot{\Psi}}$, however,
 effects until about 2.5 $\tau_H$. So that we may say that the inflaton
$\Phi$ is homogenized partially by the field $\Psi$.
Hence, to search the effects of the inhomogeneity, we had better survey
initial data without such an additional field.

Since we are working in a planar spacetime, the gravitational waves
are also exist  in this system. We calculate the Weyl tensor and study
its evolutions. In Fig. 4.6, the maximum value of
 $\sqrt{C_{\mu \nu \rho \sigma}
C^{\mu \nu \rho \sigma}}/\Lambda_{\rm eff}$ on each hypersurface for
various $\lambda$ is drawn.  If $\lambda$ is short enough, then the
initial Weyl invariant become large. Such a deviation, however,
soon homogenized by the expansion of the
universe, which results are agree with those in Section 3.

\subsection{Evolutions of the inhomogeneous scalar field}

Next, we show the time evolution of the initial
 data without introducing a massless scalar field.
This setting makes clear the effects of inhomogeneous scalar
field, since initial expansion rate are not uniform over the spacetime.

To show the difference with the previous model, we set the same
chaotic potential  with $m^2=0.01$, and the same
 initial scalar distribution form $\Phi_{\rm initial}$.
To define a unit,  we set the effective cosmological constant as
${\tilde \Lambda}_{\rm eff} \equiv  {1 \over 2} m^2 \Phi_0^2$, where
$\Phi_0$ is the mean value of the initial scalar field
 [see eq. (\ref{cos-type-scalar})],
 and fix unit $l_H$ and $\tau_H$ using ${\tilde \Lambda}_{\rm eff}$.
We set the background spacetime is flat on the initial hypersurface,
and choose $\mbox{tr}\hat{K}$ as
\be
\mbox{tr}\hat{K}=-\sqrt{3 {\tilde \Lambda}_{\rm eff}}
(1+\tilde {\delta}_K).
\label{tracecon3}
\en
Since our present model is not isocurvature, $\tilde{\delta}_K$ must be
introduced in order to take into account the inhomogeneity effect,
and is determined by the same condition as in Sec.3.

As like the cases in the previous section,  we first prepare initial data
for large amplitude $\delta\Phi$ and the scale  $\lambda$
to be larger than a horizon scale (model
[II-$2a \sim 2c$] in Table 4.2).
Initial rate of
$\rho_{\Phi'}$ to $\rho_{\rm total}$ and each
 ${\tilde{\delta}}_K$ are shown in Table 4.2.
The gradient energy rate
${\overline{\rho_{\Phi'}} / \overline{\rho_{\rm total}}}$
are large enough to treat our problem.
The inflaton in such a case behaves like a homogeneous
field in one horizon region, and induce a sufficient expansion
if the average value of $\Phi$ is `suitably' large
 for the inflationary model.
These results are again agree with GP's results
as we expected.


We next prepare initial data for small $\lambda$ compare to the
horizon scale.
With the same reason in the previous section,
if we concentrate our attentions for the small scale deviations
$\lambda$ compare to the horizon scale, then only
 small amplitude of $\delta\Phi$ is acceptable.
We  set
$\lambda=0.2l_H$  and
change $\delta \Phi$ (model [II-$2d \sim 2h$] in Table 4.2).

A typical example of the time evolution of scalar
field is shown in Fig. 4.7.
We see the field $\Phi$ is rolling down along the chaotic
inflationary potential.
In the figure, only the configuration until 12$\tau_H$
 is drawn, but we can see
differences with that of the isocurvature model (Fig. 4.2).
Since there is no uniform expansion in the initial stage,
the scalar field is not homogenized immediately but
 has different kinetic energy at the different point,
and drags the initial distribution even at  12$\tau_H$.
As seen from Fig. 4.7,  the field seems to become homogeneous around
5$\tau_H$,
this is because the field of initially larger $\Phi$ gets large
kinetic energy and goes down the potential faster than that
from initially smaller $\Phi$.
In the last stage, we find again inhomogeneous
distribution similar to the initial one.
Although the configuration of the scalar field is inhomogeneous,
the evolution of the spacetime goes to the inflationary stage, since
the field at each point follows the usual slow rolling  scenario.
The spacetime becomes locally de Sitter universe.
We can see this by using phase diagram of this model (Fig. 4.8).
For the different initial data shows that all
 the trajectories coincide at a time a
bit after the beginning and all follow the same history
with the homogeneous case.
Compare to the isocurvature case [Fig. 4.4(a)], we find the
trajectories at the very beginning are the quite different.
This difference shows the effect of an additional massless scalar field,
i.e., the initial uniform expansion make the inflaton homogeneous at
the beginning of the time evolution
[trajectories round around the homogeneous case in Fig. 4.4(a)].

An important fact is that we did not find  a collapse
 of the spacetime even if we
set the initial inhomogeneity is large enough.
 All the initial effect of the
gradient energy term $\rho_\nabla$ on the evolution
 of the spacetime do not work on preventing an onset of inflation
in a planar universe.
If the  $\Phi_0$ is large enough to derive an inflation just as same as
that in homogeneous and isotropic spacetime, then
the planar inhomogeneous spacetime go into inflationary era.

\section{Conclusion}

We analyzed the validity of the cosmic no hair conjecture  in a
plane symmetric spacetime.
We integrated the  Einstein equations numerically
using ADM formalism to see following two effects on the evolution
of the spacetime;  strong
inhomogeneity [I] due to localized gravitational waves in the spacetime
 with positive cosmological constant and [II] due to
inhomogeneous inflaton field.

For the model [I], we see all initial inhomogeneities
 are smoothed out within
one Hubble expansion time $\tau_H$ including their collisions' effect.
So that we
conclude that the nonlinearity of the gravity do not cause to
prevent the onset of the inflation under the existence of positive
cosmological constant $\Lambda$.

For the model [II], we prepare two initial situations; isocurvature
initial data and fully inhomogeneous initial data.
In both cases,
we find that initial gradient term of the energy density of the scalar
field may not lead to collapse of the spacetime.
If the initial mean value of the inflaton field, $\Phi_0$, is large
enough to derive inflation in homogeneous and isotropic spacetime,
then we can conclude such a universe will always be homogenized and
expand sufficiently.

For the case of the inhomogeneous scale $\lambda$ is larger than the
horizon scale, our calculations  confirm  the
criterion (\ref{criterion}) by Goldwirth and Piran, and also  their
calculations for spherical spacetime.

Even for the case of small $\lambda$, we could not find
the collapsing universe.
We did not face to a formation of a naked singularity
at least in a planar universe.
This may support a cosmic censorship conjecture.
Therefore, we conclude that inflation is generic, and the usual
analysis with isotropy and homogeneity may be justified in a
plane-symmetric spacetime.
We also partially confirm
our inflationary scenario of inhomogeneous universe, which
 we mentioned in the introduction.
That is,  a large inhomogeneity does not necessarily prevent
cosmological  constant-driven expansion, instead the spacetime evolve
into a de Sitter spacetime with many small black holes induced by
gravitational collapses.

As we denoted in section 4.2, the condition of isocurvature initial data
by introducing an additional massless field may lead to the different
configuration on the evolution of the spacetime.
The critical difference did not appear in the plane symmetric spacetime,
but it is necessary to treat precisely also in a spherical universe.
We are now studying the spherically symmetric
initial data without introducing such an additional field \cite{chiba}.

 In order to clear out the generality of the inflation,
 further simulations ---for example, those
 present in cylindrical or axially symmetrical, or even more general
 spacetimes---are required \cite{3d}.

\vskip 1.0cm
{\bf Acknowledgment}

We would like to acknowledge
L. Gunnarsen,  T. Nakamura,
 K. Nakao and K. Oohara for helpful discussions.
One of the author (H.S) would like to thank to
the Yukawa Institute for Theoretical Physics,
 where a part of this work was done.
This work was supported partially by the Grant-in-Aid
for Scientific Research Fund of the Ministry of Education, Science and
Culture Nos. (04640312 and 0521801), by a Waseda University Grant for
Special Research Projects and by The Sumitomo Foundation. \par



 \newpage
\begin{center}
{\bf Table Captions}
\end{center}
 \baselineskip 15pt

\noindent
Table 3.1:  \\
\noindent
Examples of initial parameters for model [I] {\bf [case 1]}.
$a$ and $x_0$ are parameters to define ${\hat \gamma}_{ij}$ in
eq.(\ref{exp-type}),
$L$ is a periodic boundary length,
and $\delta_K$ is a constant in eq.(\ref{tracecon}).
$l(t=0)$ and ${\cal C}_{\rm max}(t=0)$ are width and amplitude,
respectively, on the initial hypersurface.
\vskip 0.5cm

\noindent
Table 3.2:  \\
\noindent
Examples of initial parameters for model [I] {\bf [case 2]}.
$\tilde{a}$ and $\tilde{x}_0$ are parameters to define ${\hat
\gamma}_{ij}$
in eq.(\ref{cos-type1}), and the rest are the same with Table 3.1.
\vskip 0.5cm

\noindent
Table 4.1:  \\
\noindent
Examples of initial parameters for the isocurvature model.
$\Phi_0, \delta\Phi, \lambda$ means average, amplitude of deviations
and scale of the deviation, respectively [in eq.(\ref{cos-type1})].
Initial ratio of gradient term
$\rho_\nabla$ to the $\rho_{\rm total}$ are also shown.
The mass term  of the potential, $m$, is set as $m^2=0.01$.
\vskip 0.5cm

\noindent
Table 4.2:  \\
\noindent
Initial parameters for the fully inhomogeneous model.
 ${\tilde{\delta}}_K$ is a constant in eq.(\ref{tracecon3}).
 The mass term  of the potential is set as $m^2=0.01$.
\vskip 2.0 cm

\newpage

\begin{center}
{\bf Tables}
\vskip 1.5 cm

\begin{tabular}{ c||c|c|c|c||c|c}
\hline \hline
&&&&&\\[-.1em]
Model & $a$ & $x_0$  & $L$ & $\delta_K$ & $l(t=0)$ &
${\cal C}_{\rm max}(t=0)$ \\[.5em]
\hline
\hline
&&&&&\\[-.1em]
I-1a &  -0.050  & 0.050 & 0.667 & 0.010 & 0.085 & 9.67 \\[.5em] \hline
&&&&&\\[-.1em]
I-1b & -0.100   & 0.050  & 0.667 & 0.007 & 0.086 & 19.86\\[.5em] \hline
&&&&&\\[-.1em]
I-1c &  -0.333  & 0.080  & 1.000 & 0.026 & 0.151 & 29.62\\[.5em] \hline
&&&&&\\[-.1em]
I-1d &  -0.500  & 0.080  & 1.000 & 0.071 & 0.161 & 49.87\\[.5em] \hline
\hline
\end{tabular}

\vskip 0.5cm
{\bf Table 3.1}
\end{center}

\vskip 2.5cm

\begin{center}

\begin{tabular}{ c||c|c|c|c||c|c}
\hline \hline
&&&&&\\[-.1em]
Model & $\tilde{a}$ & $\tilde{x}_0$  & $L$ & $\delta_K$ & $l(t=0)$ &
${\cal C}_{\rm max}(t=0)$ \\[.5em]
\hline
\hline
&&&&&\\[-.1em]
I-2a & 1.00  & 0.10 & 0.667 & 0.05 & 0.079 & 0.618 \\[.5em] \hline
&&&&&\\[-.1em]
I-2b & 1.00  & 0.05 & 0.500 & 0.10 & 0.045 & 2.052\\[.5em] \hline
&&&&&\\[-.1em]
I-2c & 2.00 & 0.05  & 0.500 & 0.20 & 0.049 & 4.234\\[.5em] \hline
&&&&&\\[-.1em]
I-2d & 3.00 & 0.05  & 0.500 & 0.30 & 0.053 &  6.517 \\[.5em] \hline
\hline
\end{tabular}

\vskip 0.5cm
{\bf Table 3.2}
\end{center}

\newpage
\vskip 2.0 cm

\begin{center}

\begin{tabular}{ c||c|c|c|c|c}
\hline \hline
&&&&\\[-.1em]
Model & ~~$\Phi_0$~~ & ~~$\delta\Phi$~~ & ~~$\lambda~~$ &
(max) $\rho_\nabla / \rho_{\rm total}$  &
 $\overline{\rho_\nabla} / \overline{\rho_{\rm total}}$ \\[.5em]
\hline \hline
&&&&&\\[-.1em]
II-1a & 5.00 & 0.75 & 1.00  & 92.5\% & 46.1\%
\\[.5em]
\hline
&&&&&\\[-.1em]
II-1b & 8.00 & 0.72 & 1.00 & 85.3\% & 42.5\%
\\[.5em]
\hline
&&&&&\\[-.1em]
II-1c & 8.00 & 0.36 & 0.50 & 85.3\% & 42.5\%
\\[.5em]
\hline
&&&&&\\[-.1em]
II-1d & 8.00 & 0.06 & 1.00  & 2.37\% & 1.18\%
\\[.5em]
\hline
&&&&&\\[-.1em]
II-1e & 8.00 & 0.06 & 0.50  & 9.47\% & 4.73\%
\\[.5em]
\hline
&&&&&\\[-.1em]
II-1f & 8.00 & 0.06 & 0.250  & 37.9\% & 18.9\%
\\[.5em]
\hline
&&&&&\\[-.1em]
II-1g & 8.00 & 0.06 & 0.167  & 85.2\% & 42.5\%
\\[.5em]
\hline
\hline
\end{tabular}

\vskip 0.5cm
{\bf Table 4.1}
\end{center}

\vskip 2.5cm

\begin{center}

\begin{tabular}{ c||c|c|c|c|c}
\hline \hline
&&&&&\\[-.1em]
Model & ~~$\Phi_0$~~ & ~~$\delta\Phi$~~ & ~~$\lambda~~$
& ~~${\tilde{\delta}}_K$~~
 & ${\overline{\rho_{\Phi'}} / \overline{\rho_{\rm total}}}$
 \\[.5em]
\hline \hline
&&&&&\\[-.1em]
II$-2a$ & 8.00 & 1.50  & 1.000  & 0.625  & 99.4\%
\\[.5em]
\hline
&&&&&\\[-.1em]
II$-2b$ & 8.00 & 2.40   & 5.000  & 1.200  & 94.1\%
\\[.5em]
\hline
&&&&&\\[-.1em]
II$-2c$ & 8.00 & 3.60   & 10.00  & 2.400  & 89.5\%
\\[.5em]
\hline
&&&&&\\[-.1em]
II$-2d$ & 8.00 & 0.03 & 0.200 & 0.005 & 86.6\%
\\[.5em]
\hline
&&&&&\\[-.1em]
II$-2e$ & 8.00 & 0.06 & 0.200 & 0.010& 96.3\%
\\[.5em]
\hline
&&&&&\\[-.1em]
II$-2f$ & 8.00 & 0.18 & 0.200 & 0.045 & 99.6\%
\\[.5em]
\hline
&&&&&\\[-.1em]
II$-2g$ & 8.00 &  0.30 & 0.200 & 0.100 & 99.8\%
\\[.5em]
\hline
&&&&&\\[-.1em]
II$-2h$ & 8.00 &  0.60 & 0.200 & 0.250 & 99.9\%
\\[.5em]
\hline
\hline
\end{tabular}

\vskip 0.5cm
{\bf Table 4.2}
\end{center}

\newpage
\vskip 1.0cm
\begin{center}
{\bf Figure Captions}
\end{center}
\vskip 0.25cm
 \baselineskip 15pt

\noindent
Figure 2.1:  \\
\noindent
The accuracy of the Hamiltonian constraint equation.
The solid line and the dotted line show
the maximum error $Error(t)$ and the averaged error, respectively,
on the hypersurface at every time step
for the calculation of the inhomogeneous scalar field (the case of
model [II-2f] in Section 4.2).

\vskip 0.5cm

\noindent
Figure 3.1:  \\
\noindent
An example of the initial configuration for the case of a pulse-like
distortion ([case 1]).
This is the case of model [I-1a] in Table 3.1.
In (a), the conformal factor $\phi$ in equation
(\ref{initialconf}) is shown.  In (b), the ``curvature" ${\cal
C}(t,x)$ in (\ref{curvature}).
\vskip 0.5cm

\noindent
Figure 3.2:  \\
\noindent
Time evolutions of propagating plane waves.
(a) ``curvature" ${\cal C}(t,x)$ [eq.(\ref{curvature})] and
(b) ``super-energy" ${\cal H}(t,x)$ [eq.(\ref{superenergy})] are shown
for the initial data shown in Fig. 3.1.

\vskip 0.5cm

\noindent
Figure 3.3:  \\
\noindent
The maximal value of
${\cal I}(t,x)\equiv { \sqrt{R_{\mu \nu \rho \sigma}
R^{\mu \nu \rho \sigma}}
/ \Lambda}$ (solid line), and
${\cal C}(t,x)\equiv { \sqrt{~^{(3)\!}R_{ijkl}~^{(3)\!}R^{ijkl}}
/ \Lambda}$ (dotted line) on each hypersurface $\Sigma(t)$ are shown for
the same data with Fig. 3.1.
We find ${\cal I}(t,x) \rightarrow \sqrt{8/3}$ within one
Hubble expansion time
(homogeneous de Sitter spacetime) as expected.

\vskip 0.5cm
\noindent
Figure 3.4:  \\
\noindent
The time evolution  of the ``curvature"  ${\cal C}(t,x)$ resulting
from waves that are located closely. Two waves are the same form,
$l=0.10l_H$ and ${\cal C}_{\rm max}(t=0)=51.0$,
and the periodic distance
is $0.30l_H$. We see them collide  and in the collision region,
the ``curvature" seems to be superposed, but finally the spacetime is
homogenized by the expansion of the universe.

\vskip 0.5cm

\noindent
Figure 4.1:  \\
\noindent
A typical example of the time evolution of initially
inhomogeneous scalar
field (isocurvature model). The model [II-1b] in Table 4.1 is shown.
 We see the field $\Phi$ is rolling down along the chaotic
inflationary potential $V(\Phi)={1 \over 2} m^2 \Phi^2$ with $m^2=0.01$.
 Although the gradient term in the
initial state is locally dominated, we see the field $\Phi$ is
homogenized immediately.

\vskip 0.5cm

\noindent
Figure 4.2:  \\
\noindent
The same as Fig. 4.1, but of model [II-1g] in Table 4.1.

\vskip 0.5cm

\noindent
Figure 4.3:  \\
\noindent
Phase space diagram($\Phi - \overline{\Pi}$) indicating the behavior of
the scalar field at $x=0$ both for the inhomogeneous initial data
(dotted line) and the homogeneous initial data (solid line). The
former initial data is  [II-1g] in Table 4.1, while the latter
initial data is the same except $\delta\Phi=0$.  We find the
dotted line is coincide with the solid one soon after rolling
down, showing that the homogeneous scalar field scenario may work
even if we include much inhomogeneities.

\vskip 0.5cm

\noindent
Figure 4.4:  \\
\noindent
(a) Trajectories in $\Phi-\overline{\Pi}$ phase space for various
 kinetic terms' initial data for the isocurvature model.
 Only the beginning part is drawn.
The long-dashed line, long-dot-dashed
line, short-dashed line and solid line are that of
Model [II-1d, 1e, 1f, 1g] respectively.
All of them will coincide with the trajectory of the homogeneous
scalar model (the bold line) soon after slowing down the potential.

\noindent
(b) The evolution of the volume element det$\gamma_{ij}$
(maximum value on the each hypersurface) for each case.
The lines are the same with (a).
The bold line shows the case of homogeneous scalar field of both
$\Phi=8.0m_{\rm pl}$ and $\Phi=8.06 m_{\rm pl}$ (indistinguishable).
We see that the inhomogeneity of the scalar field enhances the
expansion of the spacetime.

\vskip 0.5cm

\noindent
Figure 4.5:  \\
\noindent
The proportions of the each terms in energy density  to
$\rho_{\rm total}$;
$\rho_{\dot{\Phi}}$(bold-solid line), $\rho_{\Phi'}$(bold-dotted line),
$\rho_{\dot{\Psi}}$(thin-solid line), $\rho_{\Psi'}$(thin-dotted line),
and  $V(\Phi)$(three-dot-dashed line).  (a) shows until $1.0\tau_H$,
while (b) shows until $5.0\tau_H$.

\vskip 0.5cm

\noindent
Figure 4.6:  \\
\noindent
The evolutions of the maximum Weyl invariant
$\sqrt{C_{\mu \nu \rho \sigma}
C^{\mu \nu \rho \sigma}}/\Lambda_{\rm eff}$ on the hypersurface.
\vskip 0.5cm

\noindent
Figure 4.7:  \\
\noindent
A typical example of the time evolution of initially inhomogeneous
scalar
field. The model [II-2f] in Table 4.2 is shown.
 We see the field $\Phi$ is rolling down along the chaotic
inflationary potential $V(\Phi)={1 \over 2} m^2 \Phi^2$ with $m^2=0.01$.
Compare to the Fig. 4.2, we notice the scalar field's configuration
drags its initial form even after the time evolution.

\vskip 0.5cm

\noindent
Figure 4.8:  \\
\noindent
Trajectories in $\Phi-\overline{\Pi}$ phase space for various
 kinetic terms' initial data for the fully inhomogeneous model
 (at $x=0$, where the maximum $\Phi$ exists initially).
 Only the beginning part is drawn.
The detail parameters are shown in Table 4.2.
The bold line shows the trajectory in the case of homogeneous scalar
field ($\Phi=8.0m_{pl}$).
The dotted line, dot-dashed line, dashed line and
solid line  are those of
Model  [II-2e, 2f, 2g, 2h]
respectively.  All of them will coincide to the homogeneous' line
soon after slowing
down the potential.


\vskip 2.0cm

\baselineskip .2in
\begin{thebibliography}{9}


\bibitem{inflation}
The original idea is in\\
A.H. Guth, Phys. Rev. {\bf D 23}, 347 (1981); \\
K. Sato, Mon. Not. Roy. Astron. Soc. {\bf 195}, 467 (1981).

\bibitem{nohair}
G.W. Gibbons, S.W. Hawking, Phys. Rev. {\bf D15}, 2738 (1977);
W. Boucher, G.W. Gibbons, G.T. Horowitz, Phys. Rev. {\bf D30}, 2447
(1984).
\\
A review is, {\em e.g.},\\
K. Maeda, in Fifth Marcel Grossmann Meeting, Proceedings,
 Perth, Australia, 1988, edited by D.Blair and M. Buchingham
(World Scientific, Singapore, 1989), p.145.

\bibitem{perturb}
H.A. Feldman, R.H. Brandenberger, Phys. Lett. {\bf 227 B}, 359 (1989);
J.H. Kung, R.H. Brandenberger, Phys. Rev. {\bf D 40}, 2532 (1989);
{\it ibid.}  {\bf D 42}, 1008 (1990).

\bibitem{counter-ex}
K.Sato, in the Proceedings of I.A.U. Symposium No.130,
{\em The Large Scale Structure of the Universe},
ed.by J.Audouze {\it et al.}, p.67 (IAU, 1988)


\bibitem{wald}
R.M. Wald, Phys. Rev. {\bf D 28}, 2118 (1983).

\bibitem{kitamae}
Y. Kitada and K. Maeda, Phys. Rev. {\bf D 45}, 1416 (1991);
 Class. Quantum  Grav. {\bf 10}, 703 (1993).

\bibitem{GP}
D.S. Goldwirth and T. Piran, Phys. Rev.
{\bf D 40}, 3263 (1989);
 Phys. Rev. Lett. {\bf 64}, 2852 (1990); Phys. Rep. {\bf 214},
223 (1992),
D.S. Goldwirth, Phys. Lett. {\bf B 243}, 41 (1990); Phys. Rev.
{\bf D 43}, 3204
(1991).

\bibitem{suonio}
H. Kurki-Suonio, J. Centrella, R.A. Matzner and J.R. Wilson, Phys.
Rev. {\bf D 35}, 435  (1987).

\bibitem{weakgw}
T. Piran, in {\it The Early Universe},
ed. by W. Unruh, (Reidel, Dordrecht, 1986); \\
K. Nakao, T. Nakamura, K. Oohara and  K. Maeda, Phys. Rev. {\bf D 43},
1788 (1991).

\bibitem{stronggw}
S.M. Miyama, Prog. Theo. Phys. {\bf 65}, 894 (1981);\\
A.M. Abrahams, C.R. Evans, Phys. Rev. {\bf D 46}, R4117 (1992).

\bibitem{colwave}
P. Szekeres, Nature {\bf 228}, 1183 (1970); J. Math. Phys. {\bf 13},
286 (1972);
\\ V.A. Khan and R. Penrose, Nature {\bf 229}, 185 (1971).

\bibitem{nakaogw}
K. Nakao, K. Maeda, T. Nakamura and K. Oohara,  Phys. Rev.
{\bf D 47}, 3194  (1993).
\bibitem{nakaodust}
K. Nakao, Gen. Rel. Grav. {\bf 24}, 1069 (1992).
\bibitem{shiromizu}
K. Nakao, K. Yamamoto and K. Maeda,  Phys. Rev. {\bf D 47}, 3203
(1993);
T. Shiromizu, K. Nakao, H. Kodama and K. Maeda, {\it ibid.}
{\bf D 47}, R3099  (1993).
\bibitem{shibata}
M. Shibata, K. Nakao, T.Nakamura and K. Maeda, preprint
WU-AP/35/93, Waseda University.




\bibitem{adm1}
J.W. York, Jr.,  in {\em Sources of Gravitational Radiation},
ed. by L. Smarr, (Cambridge, 1979) ;
N. O'Murchadha and J.W. York, Jr., Phys. Rev. {\bf D 10}, 428 (1974).

\bibitem{adm3}
T. Nakamura, K. Maeda, S. Miyama and M. Sasaki, Prog. Theo. Phys.
{\bf 63}, 1229 (1980).

\bibitem{smarr-ann}
L. Smarr, Ann. N.Y. Acad. of Sci.{\bf 302}, 569 (1977).

\bibitem{cenmatz}
J. Centrella and  R.A. Matzner, Astrophys. J. {\bf 230}, 311 (1979);
J. Centrella, {\it ibid.} {\bf 241}, 875 (1980);
J. Centrella and R.A. Matzner, Phys. Rev. {\bf D 25}, 930 (1982).


\bibitem{ACM}
P. Anninos, J.M. Centrella and R.A. Matzner,  Phys. Rev. {\bf D 39},
2155 (1989);
in {\em Frontiers in Numerical Relativity}, ed. by C.R. Evans,
 L.S. Finn and
 D.W. Hobil, (Cambridge, 1989); Phys. Rev. {\bf D 43}, 1808 (1991);
 Phys. Rev. {\bf D 43}, 1825 (1991).

\bibitem{initial0}
N. O'Murchadha and J.W. York, Jr., Phys. Rev. {\bf D 10}, 428 (1974).

\bibitem{belrob}
V.D. Zakharov, {\em Gravitational Waves in Einstein's
Theory}, (Halsted Press, Jerusalem, 1973).


\bibitem{WUAP29}
H. Shinkai and K. Maeda, Phys. Rev. {\bf D 48}, 3910 (1993).

\bibitem{chiba}
T. Chiba, K. Nakao and T. Nakamura,
in preprint KUNS-1230,  to be published in Phys. Rev.{\bf D}.\\
H. Shinkai, T. Chiba, K. Nakao and T. Nakamura, in preparation.

\bibitem{3d}
See  \cite{shibata} and
H. Kurki-Suonio, P. Laguna and  R.A. Matzner,
Phys. Rev. {\bf D 48}, 3611 (1993).
In the latter paper, some runs of the time evolutions of the
inhomogeneous scalar field by a 3-d numerical code are presented.

\end{thebibliography}
\end{document}